\documentclass[journal]{IEEEtran}
\usepackage{amsmath,amsfonts,graphicx,algorithm,algcompatible,bm,balance,amssymb}
\usepackage[OT1]{fontenc} 

\algnewcommand\INPUT{\item[\textbf{Input:}]}%
\algnewcommand\OUTPUT{\item[\textbf{Output:}]}%
\algnewcommand\RETURN{\item[\textbf{Return:}]}%
\usepackage{multicol,tabularx,capt-of}
\usepackage{hhline}
\usepackage{multirow}
\usepackage{graphicx}
\usepackage{amsthm}
\usepackage{amsmath}
\usepackage{amssymb}
\usepackage{esint}
\usepackage{amsfonts}
\usepackage{cite}
\usepackage{balance}
\usepackage{caption}
\usepackage{subcaption}
\usepackage{epstopdf}
\usepackage{color}
\usepackage{tikz}
\usepackage{longtable}

\newcommand{\bb}[1]{\mathbb{#1}}
\newcommand{\ff}[1]{\mathbf{#1}}
\newcommand{\bs}[1]{\boldsymbol{#1}}
\newcommand{\ca}[1]{\mathcal{#1}}

\newcommand{\bh}{\ff{h}}
\newcommand{\bff}{\ff{f}}

\newcommand{\ffR}{\bb{R}}

\newcommand{\bu}{\ff{u}}
\newcommand{\bv}{\ff{v}}
\newcommand{\bU}{\ff{U}}

\newcommand{\bG}{\ff{G}}
\newcommand{\bQ}{\ff{Q}}
\newcommand{\bq}{\ff{q}}
\newcommand{\bw}{\ff{w}}
\newcommand{\bth}{\bs{\Theta}}
\newcommand{\bthe}{\bs{\theta}}

\DeclareMathOperator{\diag1}{diag}

\DeclareMathOperator{\re1}{\ca{R}}
\DeclareMathOperator{\im1}{\ca{I}}

\begin{document}

\title{NeuroRIS: Neuromorphic-Inspired Metasurfaces}
\author{Christos G. Tsinos, Alexandros-Apostolos A. Boulogeorgos, and Theodoros~A.~Tsiftsis
\thanks{ C. G. Tsinos is with the National and Kapodistrian University of Athens,  Greece (e-mail: ctsinos@uoa.gr). }
\thanks{A.-A. A. Boulogeorgos is with the Department of Electrical and Computer Engineering, University of Western Macedonia, 50100 Kozani, Greece (e-mail:
al.boulogeorgos@ieee.org).
}
\thanks{T. A. Tsiftsis is with the Department of Informatics \& Telecommunications, University of Thessaly, 35100 Lamia, Greece (e-mail: tsiftsis@uth.gr).
}
\thanks{The work of A.-A. A. Boulogeorgos was supported by MINOAS. The research project MINOAS is implemented in the framework of H.F.R.I call “Basic research Financing (Horizontal support of all Sciences)” under the National Recovery and Resilience Plan “Greece 2.0” funded by the European Union – NextGenerationEU (H.F.R.I. Project Number: 15857).}}

%
\maketitle

\begin{abstract}
Reconfigurable intelligent surfaces (RISs) operate similarly to electromagnetic (EM) mirrors and remarkably go beyond Snell law to generate an applicable EM environment allowing for flexible adaptation and fostering sustainability in terms of economic deployment and energy efficiency. However, the conventional RIS is controlled through high-latency field programmable gate array or micro-controller circuits usually implementing artificial neural networks (ANNs) for tuning the RIS phase array that have also very high energy requirements. Most importantly, conventional RIS are unable to function under realistic scenarios i.e, high-mobility/low-end user equipment (UE). In this paper, we benefit from the advanced computing power of neuromorphic processors and design a new type of RIS named \emph{NeuroRIS}, to supporting high mobility UEs through real time adaptation to the ever-changing wireless channel conditions. To this end, the neuromorphic processing unit tunes all the RIS meta-elements in the orders of $\rm{ns}$ for particular switching circuits e.g., varactors while exhibiting significantly low energy requirements since it is based on event-driven processing through spiking neural networks  for accurate and efficient phase-shift vector design. Numerical results show that the NeuroRIS achieves very close rate performance to a conventional RIS-based on ANNs, while requiring significantly reduced energy consumption with the latter.

\end{abstract}


\IEEEpeerreviewmaketitle
\vspace*{-0.3cm} 
\section{Introduction}
\IEEEPARstart{R}{econfigurable} intelligent surfaces (RISs) operate like electromagnetic mirrors that go beyond the conventional Snell law in order to create a beneficial propagation environment with unprecedented excellence in terms of reliability, energy and resource efficiency. However, nowadays RIS-enabled wireless systems are unable to support realistic wireless networks with mobile user equipment (UE), while ensuring low computational power consumption, due to the high latency of their control units and switching circuits~\cite{10080950}. 

Recognizing the aforementioned problem, the authors of~\cite{1236089} replaced the low-response time PIN diodes of the switching circuit with varactor diodes. Interesting, this change reduced the response time of the switching circuit from some $\rm{\mu s}$ to $\rm{ns}$~\cite{10080950}. Inspired by this, the authors of~\cite{9749219} followed a similar approach using varactor diodes and achieved continuous phase shift, which further increased RIS efficiency {in tuning to rapid variations of the wireless environment}. 

From the control unit point of view, several different computing architectures, including application-specific integrated circuit (ASIC) low-power (LP) complementary metal-oxide semiconductor (CMOS)~\cite{9514889}, ASIC CMOS~\cite{Petrou2022}, ASIC fin field-effect transistor (FinFET) CMOS~\cite{8704523}, field programmable gate array (FPGA)~\cite{9551980}, computing processing unit (CPU)~\cite{Usman2022}, and graphical processing unit (GPU)~\cite{9690144}, were employed. However, the aforementioned architectures either experience high response time, i.e., in the orders of $\rm{\mu s}$ to $\rm{ms}$, or have dramatically high power consumption that may even reach some hundreds Watts. Fortunately, a new brain-inspired architecture was very recently developed, namely neuromorphic computing architecture. According to~\cite{isik2023survey,Rajendran2019,Davies2021}, neuromoprhic computing processing can achieve significantly low response time, which is in the orders of some $\rm{ns}$, while their power consumption is in the order $\rm{mW}$.

Despite the important role that brain-inspired architectures can play when integrated in RIS as control units, to the best of the authors' knowledge no published contribution addresses the topic of designing neuromorphic computing controlled intelligent metasurfaces. A possible reason behind this is that in order to make the most out of the neuromoprhic computing architecture, the conventional methods that are used for beam design and beam tracking should be rethought and redesigned to following the spiking neural networks (SNNs) approach. 
Motivated by the practical limitations of nowadays conventional RISs, the advancements in computing architecture, as well as the need of developing a new type of methods for RIS beamforming vector design, in this letter, we present the following technical contributions: 
\begin{itemize}
    \item We introduce a new-type of RIS, named NeuroRIS, which uses varactors as the basis of the switching circuit in order to ensure continuous phase shift, and neuromophic computing processor as a control unit to importantly reduce the RIS response time, while boosting the processing energy efficiency.
    \item Building upon the NeuroRIS particularities, we formulate and solve a signal-to-noise-ratio (SNR) optimization problem that returns the optimal beamforming transmission and RIS phase shift vectors. To address the problem, an unsupervised learning-based SNN method is employed that make the most out of the NeuroRIS architecture.
    \item Finally, simulation results that highlight the superiority of the NeuroRIS in comparison with conventional RIS, in terms of computational energy efficiency and latency are presented.  
\end{itemize}

\vspace*{-0.3cm} 
\section{NeuroRIS architecture}
\begin{figure}
\centering
    \includegraphics[width=0.45\textwidth]{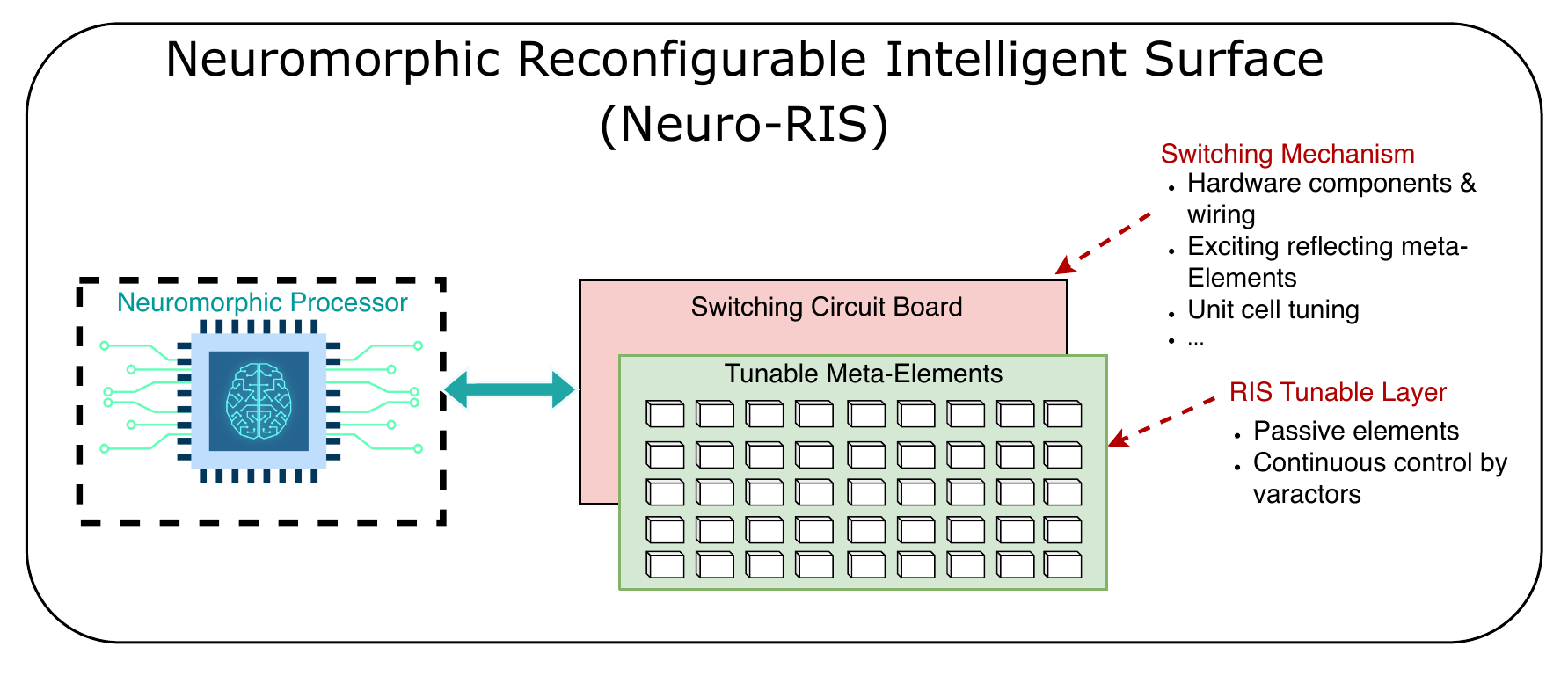}
		\caption{NeuroRIS architecture.}
  \label{Fig:NeuroRIS}
		\vspace{-6mm}
\end{figure}

The  RIS metasurface consists of $N$ unit cells that each one employ at least one  switching circuit. PIN diodes and liquid crystal switching circuits support discrete phase shift and have a response time in the orders of $\rm{\mu s}$ and $\rm{ms}$~\cite{9969596}, respectively. On the other hand, varactor diodes support both discrete and continuous phase shift~\cite{9749219} and has a response time in the orders of $\rm{ns}$. The switching circuit of the RIS is responsible for changing the state of each RIS unit. Each circuit is connected to a micro-controller that is designed according to the ASIC LP CMOS~\cite{9514889}, ASIC CMOS~\cite{Petrou2022}, ASIC FinFET CMOS~\cite{8704523}, FPGA~\cite{9551980}, CPU~\cite{Usman2022}, or GPU~\cite{9690144} principles. Excluding the ASIC FinFET CMOS, all the other ASIC architectures have a response time in the orders of $\rm{\mu s}$ and a power consumption in the orders of several decades of $\rm{mW}$~\cite{8662398,8008536}. The ASIC FinFET CMOS architectures have a similar response time with a power consumption in the range of some $\rm{W}$~\cite{8588363}. Most FPGAs support response times in the order of $\rm{\mu s}$, with power consumption that reach $10.5\,\rm{W}$~\cite{8341965}. Although,  CPUs and GPUs provide $\rm{n}s$ response times, their power consumption even reach hundreds of $W$. Note that the coherence time of millimeter wave and THz wireless channels is in the orders of some $\rm{\mu}s$~\cite{9794668}; thus, the RIS adaptation to the ever changing channel conditions should be in the orders of some $\rm{ns}$.  

As illustrated in Fig.~\ref{Fig:NeuroRIS}, to ensure a latency in the order of some $\rm{ns}$, while minimizing the power consumption of the control unit,  NeuroRIS employs brain-inspired neuromorphic computing processors for calculating its optimal phase-shifting weights through the solution of in general difficult nonconvex optimization problems.  Such processors have a significantly lower power consumption, while their operation frequency is in the order to some $\rm{GHz}$.  In more detail,  the neuromorphic processor performs all the machine learning (ML)-related calculations through SNNs.  The discrete leaky integrate and fire (LIF) neuron model proposed in \cite{LIF} is adopted for the stimulation of the membrane potential change.  For a $N_L$ layer SNN having $N_l$,  $1\leq l \leq L$ neurons per layer,  the dynamics of the $(k,l)$th neuron,  $1\leq k \leq N_L$ are given via the discretized differential equation,  given by
%
\begin{equation}
\omega_{k,l}[t+1] = \beta\omega_{k,l}[t] + I_{k,l}[t] - \beta \Theta(\omega_{k,l}[t]-\omega_{thr}), 
\label{eq:lif}
\end{equation}
where $\omega_{k,l}[t]$ is the membrane potential of the $(k,l)$th neuron,  at the simulation timestep $t$,  $1\leq t\leq T$,  $\beta$ is a decay factor,  $I_{k,l}[t] = \sum_{j=1}^{N_{l-1}}w_{j,k,l}o_{j,l-1}[t]$ is the input current calculated as the weighted sum of spikes generated by the neurons of the $(l-1)$th layer,   $w_{j,k,l} \in \mathbb{R}$ are the corresponding weights,  $o_{j,l-1[t]}$ denotes the spiking binary output of the $j$th neuron of the $(l-1)$th layer at the simulation timestep $t$,  and $\Theta(\cdot)$ is the Heaviside step function,  defined as
\vspace*{-0.2cm} 
\begin{align}
\Theta(x) = \left\{
\begin{array}{ll}
      1 & x \geq 0 \\
      0 & x < 0 \\
\end{array} 
\right. 
\vspace*{-0.2cm} 
\end{align}

The LIF neuron works as follows: When the given threshold $\omega_{t}$ is exceeded by its membrane potential,  the neuron fire a spike and the membrane potential is reset to a predetermined value $\omega_r$.  Through the simulation window $1\leq t\leq T$ the spike sequences are generated based on the timesteps that the LIF neurons fire.  Based on \eqref{eq:lif},  in the present paper we assume that $\omega_{r} = 0$ without loss of the generality.

\vspace*{-0.15cm} 
\section{System Model}
\label{sec:sys_mod}

Let us consider the downlink scenario of a system constituted by a base station (BS) equipped with $M$ antennas,  a RIS equipped with $N$ unit cells and a single antenna UE, as shown in Fig.~\ref{Fig:NeuroRIS}. The phase of the unit cells can be adaptive adjusted by a neuromorphic controller that interconnects the BS and RIS unit. The neuromorphic controller is equipped with a neuromorphic processor capable of optimizing the unit cell phases via applying learning mechanisms over SNNs.  Also, it coordinates the switching between the channel estimation and signal reflection modes.

Let us assume that there are direct transmission paths between the BS and the UE, the BS-RIS and RIS-UE. Then, the baseband received signal can be obtained~as
\vspace*{-0.1cm} 
\begin{equation}
y = (\bG\bU\bff+\bh)^T\bw s+z,
\vspace*{-0.1cm} 
\end{equation}
where  $\bw_ \in \bb{C}^{M}$ is the transmit beamforming vector,  applied at the BS side that satisfies a total transmit power constraint,  i.e.,  $\|\bw\|^2\leq P_{max}$,  where $P_{max}$ is the maximum supported transmit power,  $\ff{h} \in \bb{C}^{M}$ is the vector with the fading coefficients of the channel between the BS and the UE,   $\ff{f} \in \bb{C}^{N}$ is the channel vector between the RIS component and the UE,  $\ff{G} \in \bb{C}^{M \times N}$ channel matrix between the BS and the RIS and $z\in \bb{C}$ is a white complex Gaussian variable with variance $\sigma_z^2$. It also holds that $\bb{E}\{|s|^2\} = 1$.  Furthermore,  all the channels are assumed to be frequency flat-fading ones.  The involved channel coefficients can be estimated via different methods as shown in~\cite{}.  In addition,  $\bU = \diag1\{\bu\}$,  where $\bu=[u_1,\dots,u_N]^T \in \ca{U}^{N}$ and $\ca{U}$ denotes the set of unit-modulus complex numbers,  i.e.,  $\ca{U}=\{u \in \bb{C}|\ |u|=1 \}$.     

Following the above system model description,  the received SNR ratio at the UE is given by
\begin{align}
\gamma &= \frac{1}{\sigma_z^2}|(\bU\bth\bff+\bh)^T\bw|^2 
= \frac{1}{\sigma_z^2}|(\bQ\bu+\bh)^T\bw|^2,
\label{eq:osnr}
\end{align}
where $\bQ \in \bb{C}^{M\times N}$ is defined as $\bQ=[f_1\bG_1, f_2\bG_2,\dots,f_N\bG_N]$,  $f_n$  is the $n$th element in $\bf$ and $\bG_n$ is the $n$th column of $\bG$.    

Our aim is to derive the optimal transmit beamforming vector $\bw$ of the BS and the optimal RIS phase shifts such that the UE's receive SNR is maximized {in terms of BS transmit power and the unit-modulus constraints at RIS}.
To that end, the following optimization problem is formulated:
where the covariance matrix $\ff{Q}$ is defined as
\begin{subequations}
\begin{align}
\hspace{-200pt} \mathcal{P}_1 :\hspace{20pt} &\max_{\bw,\bu}  \frac{1}{2}|(\bQ\bu+\bh)^T\bw|^2 \\
&s.t.  \ \  \|\bw\|^2 \leq P_{max}  \\
& \ \ \ \  \ \ \bu \in \ca{U}^{N}.
\end{align}
\end{subequations}

Problem $\ca{P}_1$ is nonconvex due to the coupling of the $\bw$ and $\bu$ variables and the nonconvex unit-modulus constraints on $\bthe$.  If we assume that $\bu$ is given,  the optimal solution for $\bw$ is given by the maximum transmission ratio method,  i.e.,  $\bw_\star^T = \sqrt{P_{max}}\frac{(\bQ\bu+\bh)^H}{\|\bQ\bu+\bh\|}$.  By plugging into the objective function of $\ca{P}_1$,  the aforementioned equivalent form,  we get an equivalent form given by 
\begin{align}
 \mathcal{P}_2 :\hspace{20pt} &\max_{\bu}  \frac{1}{2}\|\bQ\bu+\bh\|^2
&s.t.  \ \ \bu \in \ca{U}^{N}.
\end{align}

Problem $\ca{P}_2$ is also nonconvex since the optimal $\bu$ still lies in the set of unit-modulus complex numbers.  Even though,  $\ca{P}_2$ is a relative simpler problem than $\ca{P}_1$ since it avoids the coupling of the $\bw$ and $\bu$ variables in,  its now convex,  objective function,  it belongs to class of NP-hard problems,  and thus it is difficult to address. To that end,  we will employ a unsupervised learning-based approach based on SNNs.

\section{Unsupervised Learning-Based Solution  SNNs}     \label{sec:snn} \vspace*{-0.1cm} 
In this section the SNN-based framework for solving $\mathcal{P}_2$ is presented.  As SNNs can only process spike sequences,  we first introduce a transformation from the feature space to the one of spike sequences  i.e.,  the coding scheme.  The corresponding inverse transformation from the spike sequence space to the one of unit-modulus beamforming vectors is then given,  i.e.,  the decoding scheme.  Then,  the loss function,  the network architecture and the the training procedure is presented in the subsequent subsections.      
\vspace*{-0.15cm} 
\subsection{Coding Scheme} \label{sec:cod}\vspace*{-0.1cm} 
From the objective function in $\mathcal{P}_2$,  one may see that the CSI related matrix $\bQ$ and CSI vector $\bh$ are suitable for constructing the feature vector for the considered SNN.  To that end, we define the feature space vector $\bv \in \ffR^{2(NM+M)}$ given by
\begin{equation}
\bv = [\re1\{\bq^T\},\im1\{\bq^T\},\re1\{\bh^T\},\im1\{\bh^T\} ]^T,
\label{eq:feat}
\end{equation}
where $\bq = \text{vec}\{\bQ\}$.  

Now the input data in $\bv$ should be transformed to spike sequences prior fed to the SNN via a coding scheme.  Among the different tested coding schemes in the literature we opt for the rate coding one due to its simplicity and robustness to noise \cite{LIF}.  At first, the entries of $\bv$ are mapped to $[0,1]$ by applying a non-linear transformation through the sigmoid function $\tilde{v}_l = 1/(1+\exp(v_l))$,  $1\leq l \leq 2(NM+M)$ and $v_l$ is the $l$th entry of $\bv$.  Then,  spike generator based on the Bernoulli distribution is employed.  In more detail,  the spiking sequence entry $r_{lt},  1\leq t \leq T$ that corresponds to the feature entry $\tilde{v}_l ,  1\leq l \leq L$ presents a spike event at time index $j$ ($r_{lt} = 1$) based on the probability
\begin{equation}
P(r_{lj} = 1) =  \tilde{v}_l = 1 - P(r_{lj} = 0).
\label{eq:spik_prob} 
\end{equation}
That is the probability of generating a spike event is proportional to the values of $\tilde{v}_l$.  
\vspace*{-0.1cm} 
\subsection{Decoding Scheme} \label{sec:dec}
The generated spiking sequence of $T$ time-steps is fed as input to the SNN and excites the neurons of the network resulting in a spiking sequence at the output.  The output spike sequence is then converted in real numbers that lie in $[0,1]$ by counting mean fire rating rates of the output layer,  that is
\begin{equation}
\tilde{\bthe} = \frac{1}{T}\sum_{j=1}^T\mathbf{o}_K(t),
\label{eq:dec}
\end{equation}
where  $\mathbf{o}_K$ is the $N\times 1$ output spike vector of the SNN's last layer at time index.  Note that the output layer is of size $N$ and each one of the output neurons correspond to the phase shift for the $n$th element in the RIS-array.  Since,  the entries of $\tilde{\bthe}$ lie in $[0,1]$,   a linear transformation is applied in order to calculate the actual phases that lie in $[0,2\pi]$,  i.e.,
  %
$\bthe = 2\pi\tilde{\bthe}$.
%
Then,  the beamforming vector entries are calculated as 
\begin{equation}
 \hat{u}_n =  e^{j \theta_n} =  \cos(\theta_n)+j\sin(\theta_n),
 \label{eq:dec3}
 \end{equation}
where $\hat{u}_n$ and $\theta_n$ are the $n$th entries of $\hat{\bu}$ and $\bthe$,  respectively,  $\hat{\bu}$ is the $N\times 1$ vector of the estimated RIS beamforming coefficients and $1\leq n \leq N$.  It is evident that by carefully setting the loss function and the training procedure,  the SNN will be able to calculate RIS beamforming vectors $\bu$ that solve~$\ca{P}_2$.  
\vspace*{-0.15cm} 
\subsection{Loss Function} \label{sec:loss}
The loss function used to train the network is based on problem  $\ca{P}_2$'s objective function,  i.e.,  
\begin{align}
\text{L} = -\frac{1}{2K}\sum_{k=1}^K\left\|\tilde{\bQ}_k
\tilde{\bu}_k+\tilde{\bh}_k\right\|^2,
\label{eq:loss_1}
\end{align}
where $K$ is the batch size,  $\tilde{\bQ}_k = \left[ {\begin{array}{cc}
   \re1\{\bQ_k\} & -\im1\{\bQ_k\} \\
   \im1\{\bQ_k\} & \re1\{\bQ_k\} \\
  \end{array} } \right]$,  $\tilde{\bu}_k = [ \re1\{\hat{\bu}_k\},  \im1\{\hat{\bu}_k\} ]^T$,  $\tilde{\bh}_k = [\re1\{\bh_k\},\im1\{\bh_k\}]^T$,  $\bQ_k$ and $\bh_k$ are the CSI related matrices $\bQ$ and $\bh$ for the $k$th training sample and $\hat{\bu}_k$ can be calculated as in  \eqref{eq:dec3}.  From  \label{eq:loss} and the considered decoding scheme in Sec. \ref{sec:dec},  it is clear that the SNN calculates a vector $\hat{\bu}_k$ that lies in the feasible solution set of  $\ca{P}_2$ and maximizes its objective function.      

\vspace*{-0.15cm} 
\subsection{Network Architecture} \label{sec:net}
The employed SNN architecture is shown in Fig. 2.  At first, the input feature vector $\bv$ is fed in on the encoder and the generated spike sequences are then led to the SNN network,  constituted by five layers of $32N$,  $16N$,  $8N$,  $4N$ and $N$ LIF neurons,  respectively.  Following such an approach,  it assures that the SNN network has the adequate learning capabilities as the NeuroRIS size $N$ scales.  Then, the output is used to calculate the mean fire rate via \eqref{eq:dec} that is subsequently used by the decoder to predict the optimal phase shifts and evaluate the SNN performance through the loss function in \eqref{eq:loss_1}. 

\begin{figure}
\centering
    \includegraphics[width=80pt, height=180pt]{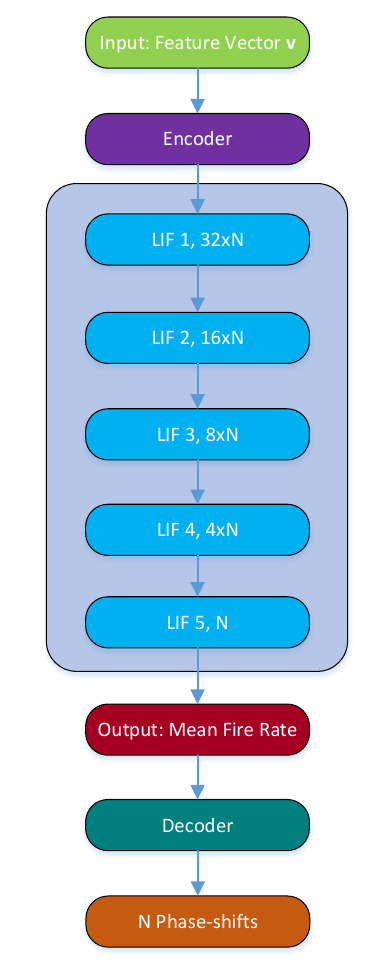} \hspace{40pt}
    	\includegraphics[width=80pt, height=180pt]{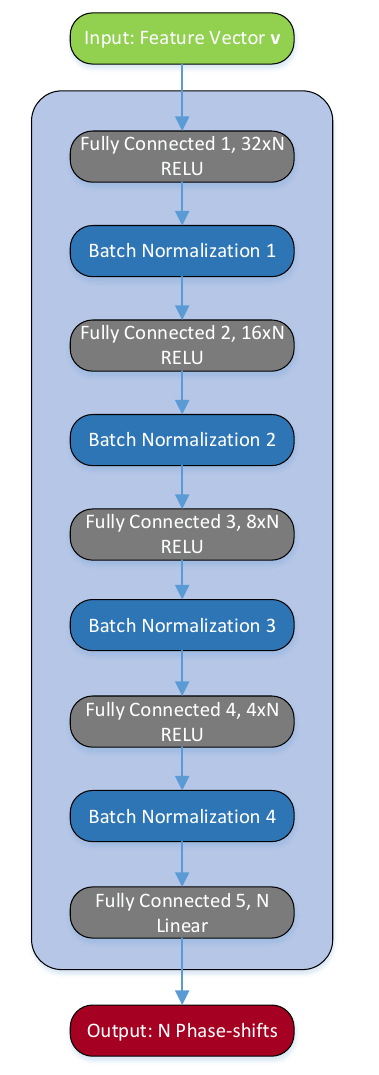}\\
    (a) \hspace{110pt} (b)
		\caption{Network Architectures for a) SNN,  and b) ANN.}
	 \label{fig:netw}
		\vspace{-7mm}
\end{figure}

\vspace*{-0.15cm} 
\subsection{Training Procedure} \label{sec:train}

The training procedure of a SNN is not a straightforward task.  Observe that from \eqref{eq:lif},  the SNN  resembles the structure of a recursive neural network. Hence,  the loss function value at time $t$ is dependent on the past values $t'$ of the learning parameters,  i.e.,  $\forall t' \leq t$.  Thus,  a generalized version of the backpropagation algorithm has to be applied for the update of the learning parameters that also captures the aforementioned temporal dependence,  i.e. ,  backpropagation through time (BPTT) \cite{slayer}-\cite{gdsnn}.  The BPTT method employs a stochastic gradient descent for updating the learning parameters.  The partial derivative of the loss function with respect to the $w_{i,k,l}$ learning parameter is given by:
\vspace*{-0.1cm} 
\begin{align}
\frac{\partial L}{ \partial w_{i,k,l}} &= \sum_{t = 0}^T\sum_{t'=0}^t \frac{\partial L[t]}{ \partial w_{i,k,l}[t']}\frac{\partial w_{i,k,l}[t']}{ \partial w_{i,k,l}[t]}  \nonumber \\
& =\sum_{t = 0}^T\sum_{t'=0}^t \frac{\partial L[t]}{\partial\Theta[t]}\frac{\partial\Theta[t]}{\partial \omega_{k,l}[t]}\frac{\partial \omega_{k,l}[t]}{\partial \omega_{k,l}[t']}\frac{\partial \omega_{k,l}[t']}{ \partial w_{i,k,l}[t']},
\label{eq:BPTT}
\end{align}
where $\Theta[t]  \triangleq \Theta(\omega_{k,l}[t]-\omega_{thres})$.  In \eqref{eq:BPTT},  the term $\frac{\partial\Theta[t]}{\partial \omega_{k,l}[t]}$ is the Dirac Delta function which is 0 everywhere apart from the point $\omega_{k,l}[t]=\omega_{thres}$ where it tends to infinity.  This results in a gradient almost always set to $0$ or saturated when the threshold is reached.  Thus,  no weight updates can be performed through the stochastic gradient descent method.  This is known as the so-called ``dead neuron problem'' \cite{LIF}.   This issue can be treated by the surrogate gradient approach that keeps the Heaviside function during the network's forward pass phase and for the backward pass phase,  $\Theta[t]$ is smoothed with the arctangent function.  Thus,  the term $\frac{\partial\Theta[t]}{\partial \omega_{k,l}[t]}$ is replaced by $\frac{\partial\tilde{\Theta}[t]}{\partial \omega_{k,l}[t]} = \frac{1}{\pi+\omega_{k,l}^2[t]\pi^3}.$
%
%
By using the term $\frac{\partial\tilde{\Theta}[t]}{\partial \omega_{k,l}[t]}$ in place of $\frac{\partial\Theta[t]}{\partial \omega_{k,l}[t]}$ during the BPTT procedure,  the dead neuron problem is avoided while the SNN's performance is not affected since neural networks are in general robust to such approximations.  The weight update procedure is iteratively executed for a predetermined number of epochs per batch of data.  Since the loss function in \eqref{eq:loss_1} is directly used on the training procedure for evaluating the SNN's  performance without generating labeled data,  the described learning method is an unsupervised one. 

\vspace*{-0.2cm} 
\section{Numerical Results}
\label{sec:num_res}

This section focuses on presenting numerical results that quantify the efficiency of the Neuro-RIS and compares its energy and data-rate performance to the corresponding performance of a conventional RIS.  In more detail,  the scenario in \cite{reg1} is considered where all the channel coefficients follow the Rayleigh model with path loss given by $20.4\,\log_{10}(d/d_{ref})\,\rm{dB}$, where $d$ is the distance between the transmitter and the receiver in meters and $d_{ref} = 1$.   The distances between BS-UE and RIS-UE can be calculated as $d_{BU} =\sqrt{d_0^2+d_1^2}$ and $d_{RU} =\sqrt{d_{BR}^2-d_0^2}$ where $d_{BR}$ is the distance between the BS and the RIS,  $d_1$ is the vertical distance from the UT to the horizontal connection line of the BS and the RIS and $d_0$ is the distance from the BS to the intersection.   The presented numerical results where generated with $d_{BR} = 8\,\rm{m}$ and $d_0$ and $d_1$ following the uniform distribution in $[0,8]$ and $[1,6]$,  respectively.  The Adam optimizer with a learning rate set to $10^{-5}$ and momentum parameters set to $\{0.9,0.99\}$ is employed for training the SNN \cite{adam}.  The simulation window size is set to $T = 25$ and parameter $\beta$ in \eqref{eq:lif} is set to $0.99$.  A number of $10^5$ samples are used to train the SNN with batch size equal to $K = 100$ for $30$ epochs.  The presented results are generated from $10^4$ test samples.   

\begin{figure}
\centering
    \includegraphics[width=0.23\textwidth]{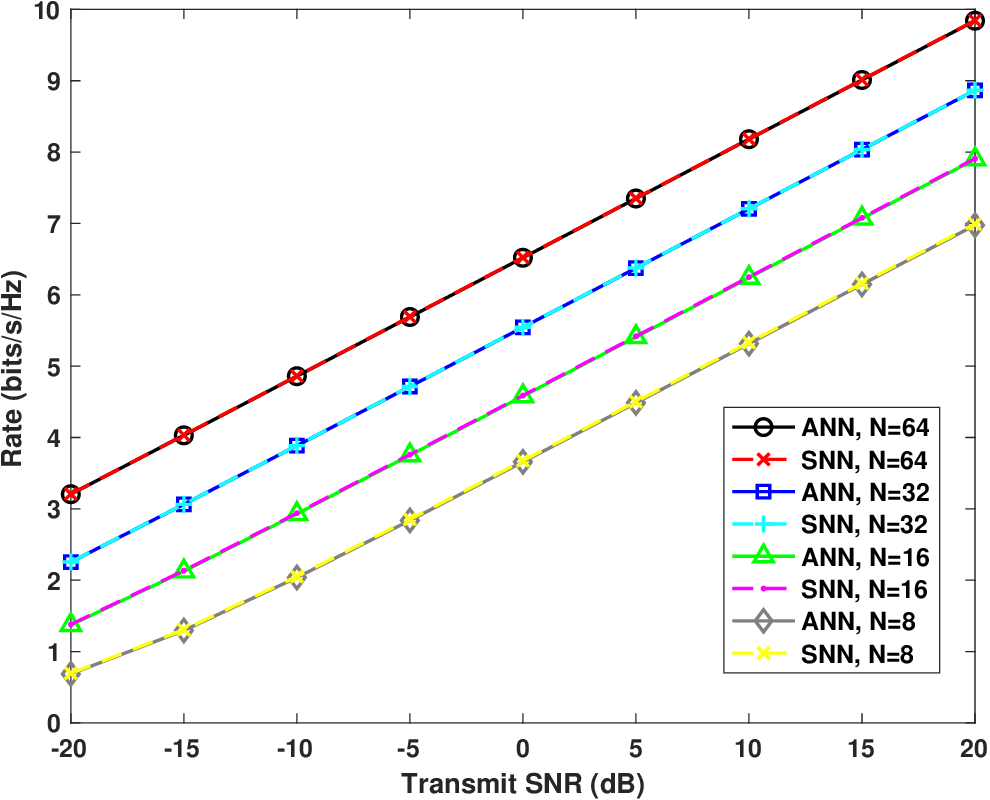} 
    	\includegraphics[width=0.23\textwidth]{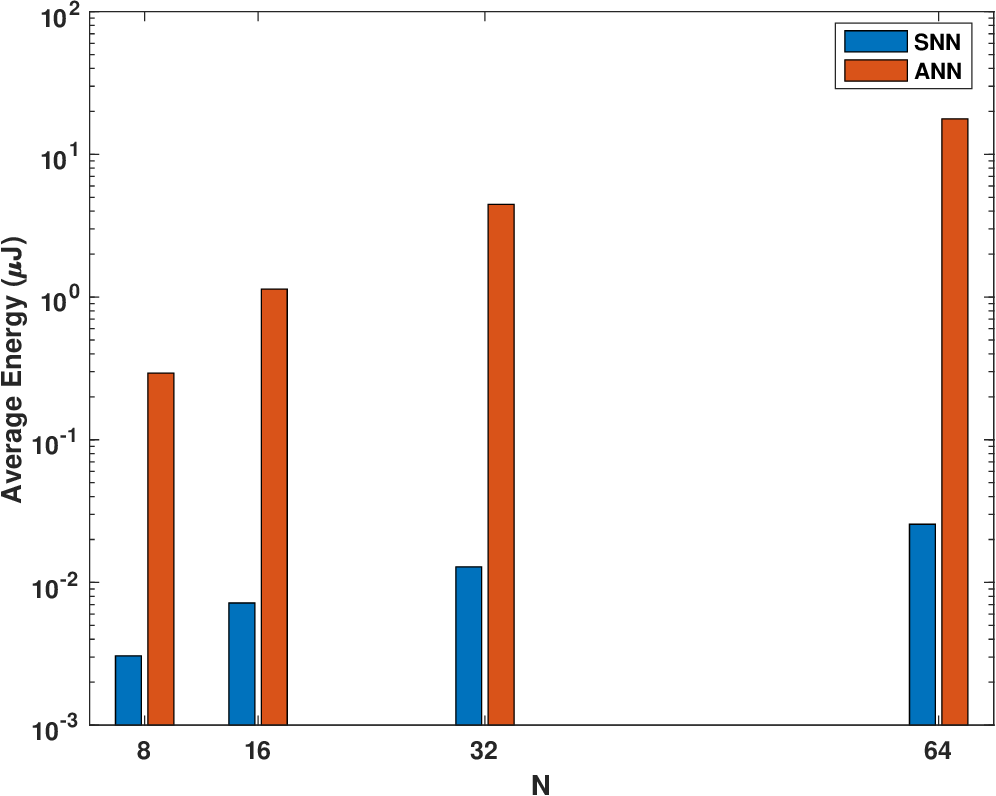}\\
    (a) \hspace{110pt} (b)
		\caption{a) Impact of the NeuroRIS size $N$ on the rate performance for a BS serving $M=2$ users, and b) Average energy consumption of ANN and SNN versus $N$. }
	 \label{fig:netw}
		\vspace{-3mm}
\end{figure}


Fig.~3a depicts the rate performance of the Neuro-RIS employing SNN to perform joint transmission and reflection beam design in terms of data rate as a function of the transmit SN for different number of RIS unit-cells.  The rate is calculated as $r = \log_2(1+\gamma)$ where $\gamma$ is calculated as in \eqref{eq:osnr}.  As benchmark,  the corresponding performance of the conventional-RIS, which uses ANN for joint  transmission and reflection beam design, is considered.  The ANN network used in the simulation follows the approach in \cite{annris} and it is shown in Fig. 2b. As expected, for given $N$ and type of RIS, as the transmit SNR increases, the data rate also increases.  For example, for $N=64$, when a NeuroRIS is employed,  the data doubles as the transmission SNR increases from $-20$ to $-5\,\rm{dB}$.  Moreover, for fixed transmission SNR and type of RIS,  as $N$ increases, the data rate increases.  For instance, in case a NeuroRIS is used, and a transmission SNR equal to $-15\,\rm{dB}$,  the data rate increases by about four times, as $N$ increases from $8$ to $64$. Finally, we observe that the NeuroRIS and conventional RIS achieve the same results in terms of data rate.      
%
Fig. 3b illustrates the average energy consumption of the RIS micro-controller as a function of $N$, assuming that the transmission SNR is equal to $0\,\rm{dB}$.  The energy consumption is calculated by multiplying the floating-point operations (FLOPS) number by their energy cost and then summing over the network's layers.  In the present paper,  we rely on the results in \cite{energy} for the energy cost per operation. There the energy cost per multiplication and addition in a $45 \,\rm{nm}$ chip operated at $0.9 \,\rm{V}$ that implements a $32$-bit architecture is given as $3.7 \,\rm{pJ}$ and $0.9 \,\rm{pJ}$,  respectively.  

As expected, for a given micro-controller, as $N$ increases, the average energy consumption increases. For NeuroRIS, as $N$ increases from $8$ to $64$, the average energy consumption increases from $3\times 10^{-3}$ to $3\times 10^{-2}\,\rm{\mu J}$. On the contrary, if a conventional RIS is used, for the same $N$ increase, the average energy consumption increases from $0.2$ to $20\,\rm{\mu J}$. Meanwhile, for the same $N$, an important energy consumption reduction is observed, when neuromorphic computing processor is used instead of a conventional micro-controller. For instance, for $N=64$, using NeuroRIS instead of conventional RIS cause an energy consumption reduction of about $3$ orders of magnitudes.  This is the case since the SNNs networks are performing mainly AC operations which consume much lower energy that the MAC operations that are the core of ANNs.  Furthermore,  due to the sparsity of the spike events,  the AC operations in SNNs are much less than the MAC ones in the ANNs.  Notice that as demonstrated in Fig.~3a, NeuroRIS and conventional RIS achieve almost identical performance for the same transmission SNR and $N$.  Hence, the use of NeuroRIS instead of conventional-RIS results to an unprecedented energy efficiency~enhancement.      


\vspace*{-0.15cm} 
\section{Conclusion}
\label{sec:conc}
In this paper, we introduced a novel RIS paradigm by exploiting the high energy efficiency of neuromorphic processors. NeuroRIS significantly outperform conventional RIS in terms of energy efficiency with same data rate performance. To highlight the important performance gains of NeuroRIS compared to conventional RISs, we formulated a joint transmission beam and RIS phase shift optimization problem that maximizes the received SNR, and we solved it by using SNNs. As a benchmark, we use the ANN corresponding solution. Our results revealed the correctness of our hypothesis, i.e., that NeuroRIS is a remarkable energy efficient solution leading to important energy saving, especially in large RIS utilizations.

\balance

\bibliographystyle{IEEEtran}
\bibliography{refs}
\end{document}